\documentclass[aps,pra,showpacs,twocolumn,superscriptaddress]{revtex4-1}
\usepackage{amssymb}
\usepackage{graphicx}
\usepackage{appendix}
\usepackage{dcolumn}
\usepackage{bm}
\usepackage{amsmath}
\usepackage{amsthm}
\usepackage{mathtools}
\usepackage{epstopdf}
\usepackage{xcolor}
\usepackage{float}
\usepackage{braket}
\usepackage[english]{babel}
\usepackage{epstopdf}
\usepackage{appendix}
\usepackage[colorlinks,linkcolor=magenta,citecolor=blue,urlcolor=blue]{hyperref}

\begin{document}

\title{Fate of Majorana zero modes by a modified real-space-Pfaffian method and mobility edges in a one-dimensional quasiperiodic lattice}
\author{Shujie Cheng}
\affiliation{Department of Physics, Zhejiang Normal University, Jinhua 321004, China}

\author{Yufei Zhu}
\affiliation{Department of Physics, University of Otago, P.O. Box 56, Dunedin 9054, New Zealand}

\author{Gao Xianlong}
\thanks{gaoxl@zjnu.edu.cn}
\affiliation{Department of Physics, Zhejiang Normal University, Jinhua 321004, China}
\author{Tong Liu}
\thanks{t6tong@njupt.edu.cn}
\affiliation{Department of Applied Physics, School of Science, Nanjing University of Posts and Telecommunications, Nanjing 210003, China}

\date{\today}

\begin{abstract}
  We aim to study a one-dimensional $p$-wave superconductor with quasiperiodic on-site potentials. A modified real-space-Pfaffian method is applied to calculate the topological invariants. We confirm that the Majorana zero mode is protected by the nontrivial topology the topological phase transition is accompanied by the energy gap closing and reopening. In addition, we numerically 
find that there are mobility edges which originate from the competition between the extended $p$-wave pairing and the localized quasi-disorder. We qualitatively analyze the influence of superconducting pairing parameters and on-site potential strength on the mobility edge. In general, our work enriches the research on the $p$-wave superconducting models with quasiperiodic potentials. 
\end{abstract}

\maketitle

\section{Introduction}
The Marjorana fermion is a  kind of particular particle, whose antiparticle is itself \cite{MF_1,MF_2,MF_3}. During the decade, the signatures of its presence have been found in 
many condensed matter systems, such as the semiconductor nanowires with strong spin-orbital couplings \cite{SOC_nano1,SOC_nano2,SOC_nano3,SOC_nano4,SOC_nano6}, 
ultracold atoms \cite{atom_1,atom_2,atom_3,atom_4}, magnetic atom chians \cite{chain_1,chain_2,chain_3,chain_4}, and the heterojunctions of normal superconductor 
and topological insulator \cite{hj_1,hj_2,hj_21,hj_3,hj_4,hj_5,hj_6,hj_7}. Because of its application prospect in topological quantum computation \cite{cpt_1,cpt_2}, 
the Majorana fermion has attracted extensive research interests \cite{atten_1,atten_2,atten_3,atten_4,atten_5}.

The Majorana fermion is theoretically proven to exist in topological superconductors with $p$-wave pairings, which appears in Majorna zero mode (MZM) and 
are located at both ends of the system, and are protected by the topology \cite{Kitaev}. As we all know, disorder will give rise to the localization phenomenon \cite{local} 
which will result in the destruction of the topological non-trivial phases in topological 
superconductors \cite{rd_1,rd_2,rd_3,rd_4,rd_5,rd_6,rd_7,rd_8}. Cai et al. \cite{cai} discussed the influence of the correlated disorder namely the quasiperiodic disorder on the MZM. 
They found that with the increase of the disorder potential, the system would undergo the transition from the topological non-trivial 
phase to the Anderson localized phase, that is, the MZM keeps robustness only to the weak disorder. Moreover, such a transition can  be characterized by the 
quench dynamics \cite{zeng} and the Kibble-Zurek machanism \cite{KZ}. Wang et al. \cite{Wang} detailedly investigated the delocalization properties of the 
topological phase where the MZM exists, and revealed that it consists of an extended phase and a critical phase. Non-Hermiticity usually brings some novel phenomena, 
such as the anomalous boundary states \cite{anomalous} and the skin effect \cite{skin_1,skin_2,skin_3}. Recent studies have shown that the MZM can appear 
in non-Hermitian $p$-wave superconductors \cite{NHTS_1,NHTS_2}, and an unconventional real-complex transition of the energies is investigated \cite{NHTS_2}.

From Kitaev's seminal work, we know that the presence and absence of the above mentioned MZM can be charactered by the Pfaffian \cite{Kitaev}. In principle, 
topological phase diagram of a generalized $p$-wave superconducting model can be further extracted by the Pfaffian method. However, due to 
the fact that this method is limited in those superconducting models possessing either homogeneous potential \cite{Kitaev} or integer-periodic potential \cite{periodic}. 
For the quasiperiodic $p$-wave superconducting models, the original Pfaffian method is no longer convenient in calculation. Therefore, in this paper, we will study the topological 
phase diagram of a $p$-wave superconducting model with a generalized potential by introducing a modified real-space-Pfaffian method. Besides, we will validate the 
efficiency of this method by other generations of Kitaev model. In addition, we note a fact that the Kitaev model with a homogeneous potential has the topological boundary 
located at $V=2t$ \cite{Kitaev} while the quasiperiodic potential possesses a wider topological boundary \cite{cai,Wang}. It is unpredictable that how the topological boundary 
changes when the homogeneous potential and the quasiperiodic potential coexist. Meanwhile, previous researches reveal that the homogeneous potential makes wave functions 
extended while the quasi-periodicity brings about the Anderson localization \cite{cai,Wang, AAmodel}. It is unknown whether there exists mobility edges when both two 
types of potentials coexist. In this paper, we will consider a special case that the potential consists of a homogeneous part and a quasiperiodic one, and make attempt to 
reveal the topological properties and mobility edges that this potential leads to. 


The rest of this paper are organized as follows.  In Sec.~\ref{S2}, we introduce the Hamiltonian of a $p$-wave superconducting model with a general potential. 
In Sec.~\ref{S3}, we first introduce the modified real-space-Pfaffian method, and then we discuss the topological properties and the mobility edges of the 
system. We make a brief summary in Sec.~\ref{S4}.

\section{model and Hamiltonian}\label{S2}
Real quantum systems are more or less affected by disorder. In this paper, we study a one-dimensional $p$-wave superconductor with quasiperiodic disordered 
on-site potentials, which is described by the following tight-binding Hamiltonian 
\begin{equation}\label{eq1}
\hat{H}=\sum^{L-1}_{n=1}\left(-t\hat{c}^\dag_{n}\hat{c}_{n+1}+\Delta\hat{c}^\dag_{n+1}\hat{c}^\dag_{n}+h.c.  \right)+\sum^{L}_{n=1}V_{n}\hat{c}^\dag_{n}\hat{c}_{n} 
\end{equation}
where $\hat{c}_{n} (\hat{c}^\dag_{n})$ denotes the fermion annihilation (creation) operator,  and $L$ is the size of the system with $n$ being the site index. The nearest 
neighbor tunneling strength $t$ and the nearest superconducting pairing parameter $\Delta$ are real constants. $t=1$ is set as the unit of energy. The quasiperiodic 
on-site potential $V_{n}$ is
\begin{equation}\label{potential}
V_{n}=\frac{V}{1-b\cos(2\pi\alpha n)},
\end{equation}
where $V$ is the potential strength, $b \in (0,1)$ is the modulation parameter, and $\alpha=(\sqrt{5}-1)/2$ is the incommensurate modulation frequency.  When $b=0$, 
the model goes back to the Kitaev model \cite{Kitaev}, where $V=2t$ is the phase transition point, namely the topological boundary of the topological 
non-trivial phase ($V<2t$) and the topological trivial one ($V>2t$). The potential can be understood as the superposition of a homogeneous potential and incommensurate potentials with different frequencies  
\begin{equation}
\begin{aligned}
V_{n}&=\frac{V}{\tanh\beta}\cdot\frac{\sinh\beta}{\cosh\beta-\cos(2\pi\alpha n)}\\
&=\frac{V}{\tanh\beta}\sum^{\infty}_{r=-\infty}e^{-\beta|r|}e^{ir(2\pi\alpha n)}\\
&=\frac{V}{\tanh\beta}\left[1+2\sum^{\infty}_{r=1}e^{-\beta r}\cos\left[r\left(2\pi\alpha n\right)\right]\right],
\end{aligned}
\end{equation}
where $\cosh\beta=1/b$ is the constraint condition. The parameter $b$ controls the number of the summation terms. When $b$ is small, $V_{n}$ can be truncated into the 
summation with finite terms. As discussed before, the homogeneous potential manifests the topological boundary of $p$-wave superconductor located at $V=2t$, whereas 
Cai et al. \cite{cai} and Wang et al. \cite{Wang} show that the considered quasi-periodicity broadens the topological boundary. For our model, it is unknown how the fate of 
topological boundary changes when both two types of potentials coexist. In addition, we know that the homogeneous potentials keep the wave functions extended while 
the incommensurate potentials lead to the Anderson localization \cite{cai,Wang,AAmodel}. Whether there exist mobility edges is still unknown when both two types of potentials coexist. 
In the next section, we will investigate these two aspects. 

In the particle-hole picture, the Hamiltonian is diagonalized. In order to obtain the full energy spectrum, we shall introduce the Bogoliubov-de Gennes (BdG) transformation
\begin{equation}
\hat{\xi}^\dag_{j}=\sum^{L}_{n=1}\left[u_{j,n}\hat{c}^\dag_{n}+v_{j,n}\hat{c}_{n}\right],
\end{equation}
where $j$ ranges from $1$ to $L$ and the components $u_{j,n}$ and $v_{j,n}$ are real numbers. Thus, the Hamiltonian in Eq.~(\ref{eq1}) is diagonalized as 
\begin{equation}
\hat{H}=\sum^{L}_{j=1}E_{j}(\hat{\xi}_{j}^\dag\hat{\xi}_{j}-\frac{1}{2}),
\end{equation}
 where $E_{j}$ is the eigenenergy which can be determined by the following BdG equations
\begin{equation}
\left\{
\begin{aligned}
-t(u_{n-1}+u_{n+1})+\Delta(v_{n-1}-v_{n+1})+V_{n}u_{n}&=E_{j}u_{n},\\
t(v_{n-1}+v_{n+1})+\Delta(u_{n+1}-u_{n-1})-V_{n}v_{n}&=E_{j}v_{n}.
\end{aligned}
\right.
\end{equation}
Furthermore, we represent the wave function as the following form
\begin{equation}
\ket{\psi_{j}}=(u_{j,1},v_{j,1},u_{j,2},v_{j,2},\cdots,u_{j,L},v_{j,L})^{T},
\end{equation}
then, according to the BdG equations, we have the following BdG matrix
\begin{equation}
\mathcal{H}=\left(
\begin{array}{ccccccc} 
A_{1} & B & 0 & \cdots & \cdots & \cdots & C \\
B^\dag & A_{2} & B & 0 & \cdots & \cdots & 0 \\
0 & B^\dag & A_{3} & B & 0 & \cdots & 0 \\
\vdots & \ddots & \ddots & \ddots & \ddots & \ddots & \vdots \\
0 &\cdots & 0 & B^\dag & A_{L-2} & B & 0\\
0 &\cdots & \cdots & 0 & B^\dag & A_{L-1} & B \\
C^{\dag} & \cdots & \cdots & \cdots & 0 & B^\dag & A_{L} 
\end{array}
\right), 
\end{equation}
where 
$
A_{j}=\left(
\begin{array}{cc}
V_{j} & 0 \\
0 & -V_{j}
\end{array}
\right)$ and  
$B=\left(
\begin{array}{cc}
-t & -\Delta\\
\Delta & t
\end{array}
\right)
$; $C$ is a null matrix when considering open boundary condition (OBC) and $C=\left(
\begin{array}{cc}
-t & \Delta \\
-\Delta & t
\end{array}
\right)$ for the periodic boundary condition \cite{Wang}. Intuitively, $\mathcal{H}$ is a $2L \times 2L$ matrix. By using the Schmidt orthogonal decomposition 
method to diagonalize the BdG matrix, we can acquire the full energy spectrum $E_{j}$ and the associated wave functions $\ket{\psi_{j}}$ directly. 
These strategies are favorable for studying the topological properties, such as the MZM and energy gap,  as well as the mobility edges. These 
investigations will be presented in the following section.

\section{results and discussions}\label{S3}
The topological property of the system is directly characterized by a topological invariant. According to Kitaev's work, the Hamiltonian 
in Eq.~(\ref{eq1}) can be  expanded in terms of Majorana operators as 
\begin{equation}
\hat{H}=\frac{i}{4}\sum^{2L}_{\ell,m}h_{\ell m}\lambda_{\ell}\lambda_{m},
\end{equation}
where $h_{lm}$ is real antisymmetric matrix, satisfying
\begin{equation}
h^{*}_{\ell m}=h_{\ell m}=-h_{m\ell}, 
\end{equation} 
and $\lambda_{\ell}$ is the Majorana operator with $\{\lambda_{\ell},\lambda_{m}\}=2\delta_{\ell m}$,  which is defined as 
\begin{equation}
\begin{aligned}
\lambda_{2n-1}&\equiv\hat{c}^\dag_{2n-1}+\hat{c}_{2n-1}=\lambda^{A}_{n},\\
\lambda_{2n}&\equiv i(\hat{c}^\dag_{2n}-\hat{c}_{2n})=\lambda^{B}_{n}.
\end{aligned}
\end{equation}
Accordingly, under PBC, the represented Hamiltonian is 
\begin{equation}
\begin{aligned}
&\hat{H}=\frac{i}{4}\left[\sum^{L-1}_{n=1}(\Delta-t)\lambda^A_{n}\lambda^{B}_{n+1}+(\Delta+t)\lambda^{B}_{n}\lambda^{A}_{n+1}\right.\\
&\left.+\sum^{L}_{n=1}V_{n}\lambda^{A}_{n}\lambda^{B}_{n}+ (\Delta-t)\lambda^{A}_{L}\lambda^{B}_{1}-(t+\Delta)\lambda^{A}_{1}\lambda^{B}_{L}-h.c.\right].
\end{aligned}
\end{equation}

For an antisymmetric matrix,  its Pfaffian is defined as 
\begin{equation}\label{Pf}
{\rm Pf}(h)=\frac{1}{2^L L!}\sum_{\tau\in S_{2L} }sgn(\tau)h_{\tau(1),\tau(2)}\cdots h_{\tau(2L-1),\tau(2L)}, 
\end{equation} 
where $S_{2L}$ denotes a series of permutations on these $2L$ elements with $sgn(\tau)$ being the sign of permutation.  
With the Pfaffian of the system, then the topological invariant $M$ can be defined as \cite{Kitaev, cai}
\begin{equation}\label{M}
M=sgn({\rm Pf}(h)).
\end{equation}

Although we known that we can calculate the Pfaffian to obtain the topological invariant of the $p$-wave superconducting system, 
Eq.~(\ref{Pf}) is appropriate for some special $p$-wave superconducting systems, such as the systems either with homogeneous potential \cite{Kitaev} 
or with integer-periodic potential \cite{periodic}. But for the quasiperiodic case, it is difficult to deal with the perturbation group operation directly. Therefore, we 
propose a modified real-space-Pfaffian method to conveniently calculate the topological invariant. This method require us to make a Schur decomposition \cite{schur} 
on the real anti-symmetric matrix $h$ 
\begin{equation}
h=UDU^{T},
\end{equation}
where $U$ is a unitary matrix, and $D$ is an anti-symmetric tridiagonal matrix. Thus, the Pfaffian of $h$ is redefined as 
\begin{equation}\label{Pf_new}
{\rm Pf}(h)=Det(U){\rm Pf}(D),
\end{equation}
$Det(U)$ is the determinant of the unitary matrix $U$. In practice, we can use Eqs.~(\ref{M}) and (\ref{Pf_new}) to obtain the topological phase diagram, for the 
reason that $Det(U)$ and ${\rm Pf}(D)$ are numerically available (see details in the Appendix). 

\begin{figure}[H]
\centering
\includegraphics[width=0.5\textwidth]{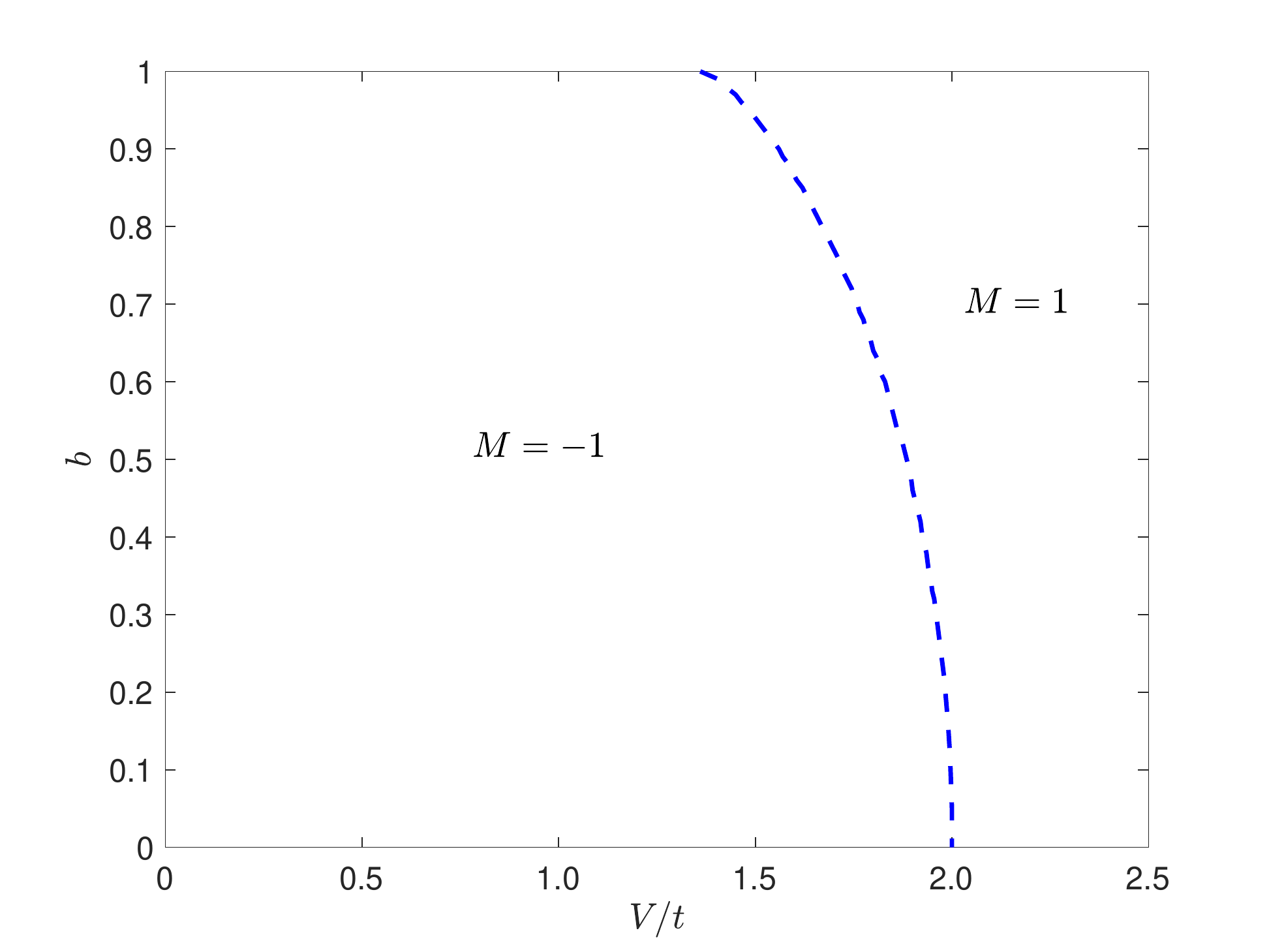}\\
\caption{(Color Online) $b-V/t$ topological phase diagram for systems with $\alpha=(\sqrt{5}-1)/2$, $\Delta=0.5t$. $M=-1$ corresponds 
to the topological non-trivial phase, and $M=1$ corresponds to the topological trivial phase. The blue dashed line denotes the phase boundary.}\label{f1}
\end{figure}

We calculate the Pfaffian by the proposed modified real-space-Pfaffian method, and naturally obtain the topological phase diagram of the system, which is 
presented in Fig.~\ref{f1}.  The diagram shows that $M=-1$ corresponds to the topological non-trivial phase, whereas $M=1$ corresponds 
to the topological trivial phase and the blue dashed line denotes the numerically obtained phase boundary. We know that when $b=0$, our 
model goes back to the Kitaev model, whose transition is located at $V=2t$ \cite{Kitaev}. We notice that when $b$ is taken at small value, the phase 
transition is almost the same with that of the Kitaev model. This implies that our model is robust against the disordered perturbations.  When $b$ increases, 
the phase boundary bends in the direction of the decreasing $V$. This phenomenon is the direct result of the enhanced disorder effect, which compress 
the topological non-trivial region. It is to say that we can not only realize the topological phase transition by adjusting the potential strength $V$, but also manipulate 
the phase transition by controlling the strength of the disorder which is determined by the the parameter $b$ in our system. 

\begin{figure}
\centering
\includegraphics[width=0.45\textwidth]{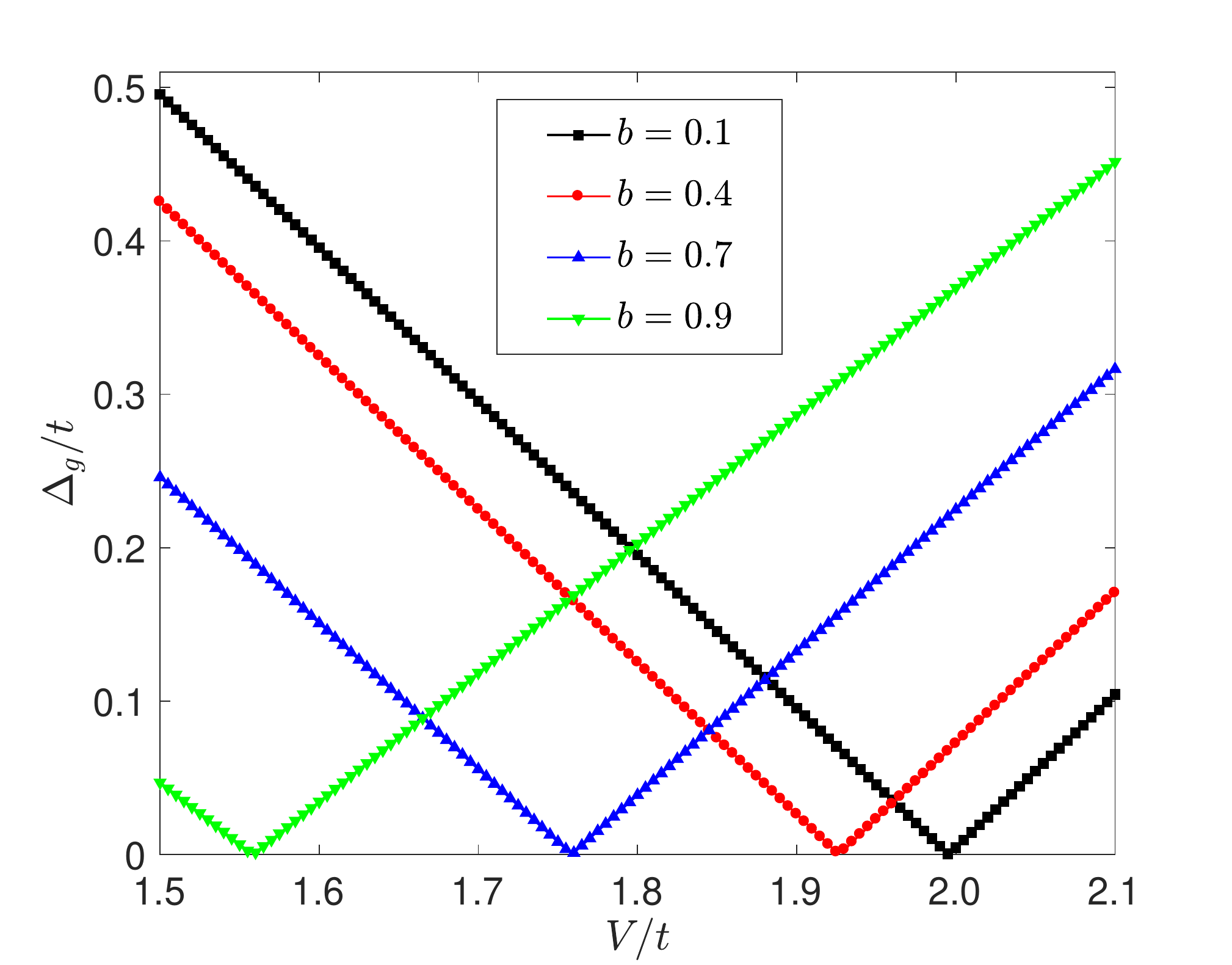}\\
\caption{(Color Online) The energy gap $\Delta_{g}$ versus $V$ with various $b$. Other involved parameters are $\alpha=(\sqrt{5}-1)/2$, 
$\Delta=0.5t$, and $L=1000$.}\label{f2}
\end{figure}
 
In topological systems, the topological phase transition is accompanied with the gap closing and reopening. 
Figure \ref{f2} plots the energy gap $\Delta_{g}$ as a function of the potential strength $V$ with 
various $b$. The $\Delta_{g}$ is defined as the difference of the $(L+1)$-th energy level and the $L$-th energy level under periodic boundary 
condition, i.e. $\Delta_{g}=E_{L+1}-E_{L}$. It is readily seen that when the topological phase transition happens, the gap undergoes closing and reopening. 
Different from the quasiperiodic case \cite{cai,Wang}, there appears a wider gap when system in the topological trivial phase. This phenomenon occurs because 
the uniform potential energy has more impact on the energy gap than that of quasi-periodic disordered perturbations \cite{Kitaev,cai}.  Moreover, when $b$ is small, 
the gap closing point is almost at $V= 2t$. As $b$ increases, the gap closing point moves towards the direction of 
the decreasing $V$. This feature is in accord with the topological boundary in Fig.~\ref{f1}.

\begin{figure}[H]
\centering
\includegraphics[width=0.5\textwidth]{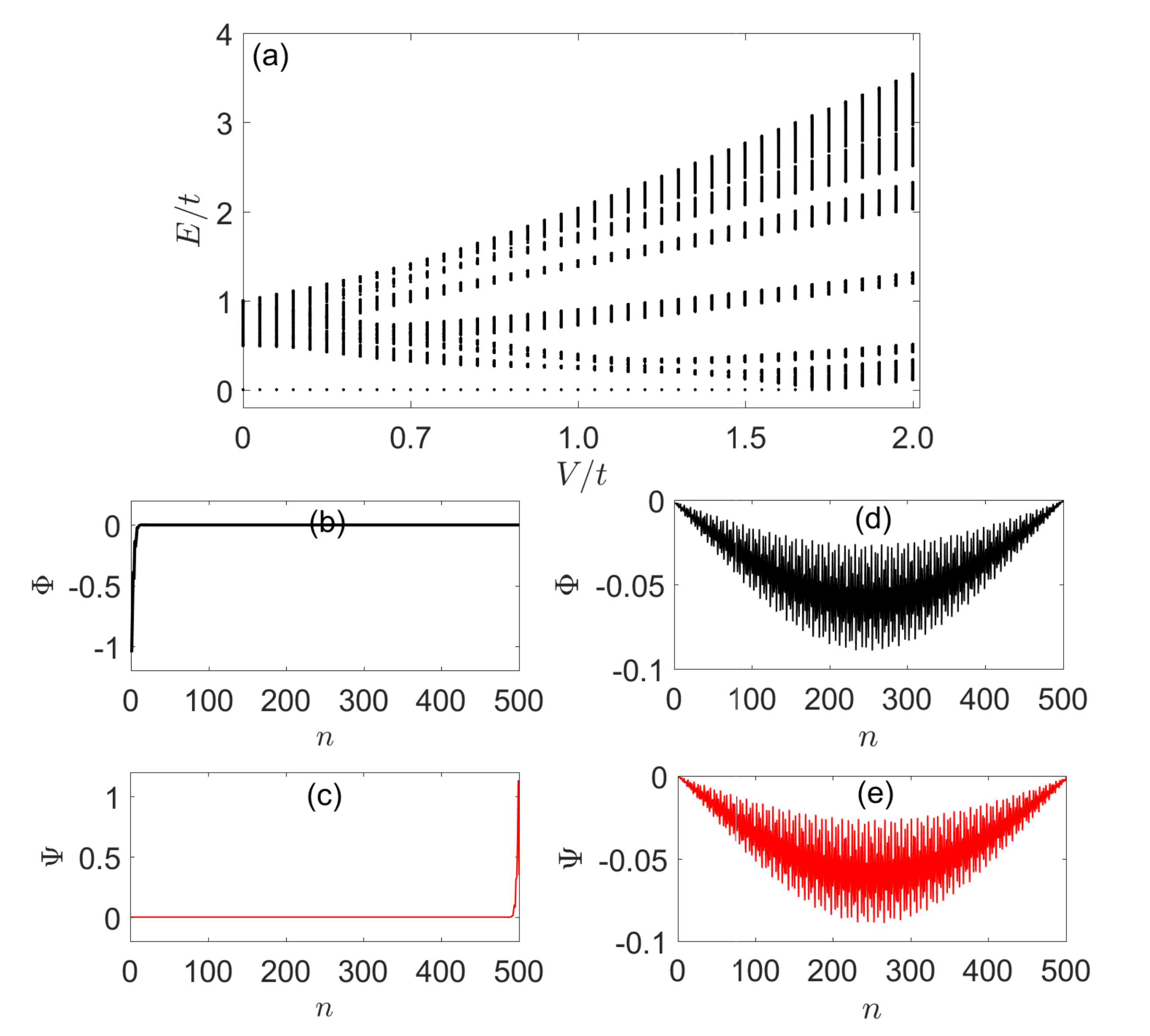}\\
\caption{(Color Online) (a) Excitation energy spectrum of the system under the open boundary condition. (b) and (c) ((d) and (e)) are respectively the spatial 
distributions of $\Phi$ and $\Psi$ for the lowest excitation state with $V=1.5t$ ($V=2t$). Other involved parameters are $L=500$, $b=0.7$, $\alpha=(\sqrt{5}-1)/2$, 
and $\Delta=0.5t$. }\label{f3}
\end{figure}

The topological non-trivial phase implies the presence of the MZMs. Figure \ref{f3}(a) shows the excitation energy spectrum 
as a function of the potential strength $V$ under the open boundary condition. The spectrum reflects that the MZMs are only 
located in the topological non-trivial phase. To see the distributions of zero energy states, we rewrite the BdG operators as 
\begin{equation}
\eta^\dag_{j}=\frac{1}{2}\sum^{L}_{n=1}[\Phi_{j,n}\lambda^{A}_{n}-i\Psi_{j,n}\lambda^{B}_{n}],
\end{equation} 
where $\Phi_{j,n}=(u_{j,n}+v_{j,n})$ and $\Psi_{j,n}=(u_{j,n}-v_{j,n})$.

Figures \ref{f3}(b) and \ref{f3}(c) respectively plot the spatial distributions $\Phi$ and $\Psi$ for the lowest excitation state with $V=1.5$. Figures \ref{f3}(d) and 
\ref{f3}(e) are distributions for the lowest excitation state with $V=2t$. When $V=1.5t$, we know that the system is in the topological non-trivial phase, so the 
lowest excitation state is the Majorana zero energy state. As the figures show, the distributions of corresponding $\Phi$ and $\Psi$ are located at ends of 
the system, reflecting the bulk-boundary correspondence. The contrary consequence is that when $V=2t$, the system is in the topological trivial phase and the 
corresponding $\Phi$ and $\Psi$ distribute in the bulk of the system. It is interpreted that the system is in the 
topological trivial phase and the lowest excitation state is no longer the Majorana edge state but the bulk state.  

We note that in the topological trivial phase, the spatial distributions of $\Phi$ and $\Psi$ for the lowest excitation state are no longer localized in the bulk but 
expand throughout the whole system, presenting an extended state. We are aware that such a phenomenon in this quasiperiodic superconducting system 
has relevance with the mobility edge instead of the Anderson localization \cite{cai,Wang}. The localization-delocalization property can be characterized by 
the inverse participation ratio (\rm{IPR}). For a given normalized wave function, the associated \rm{IPR} is defined as 
\begin{equation}
{\rm{IPR}}_{j}=\sum^{L}_{n=1}\left(|u_{j,n}|^4+|v_{j,n}|^4\right).
\end{equation}

\begin{figure}
\centering
\includegraphics[width=0.5\textwidth]{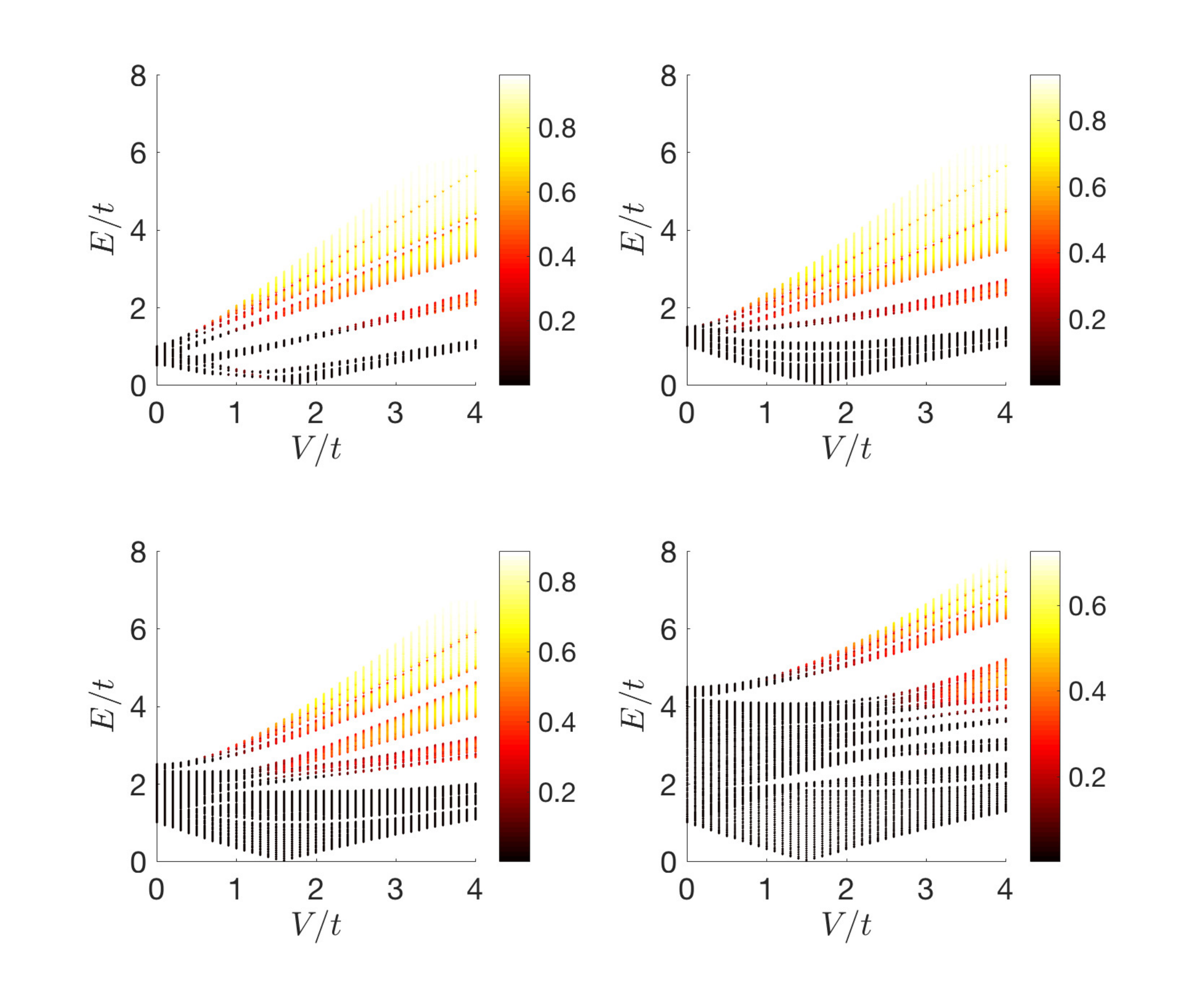}\\
\caption{(Color Online) The excitation spectrum and IPR as a function of $V$ with $\Delta=0.5t$ in (a), $\Delta=1.5t$ in (b), $\Delta=2.5t$ in (c), and 
$\Delta=4.5t$ in (d).  Other involved parameters are $b=0.7$, $\alpha=(\sqrt{5}-1)/2$, and $L=500$.}\label{f4}
\end{figure}

It is well know that for an extended wave function, the \rm{IPR} scales like $L^{-1}$ and it approaches $1$ for a localized wave function. We consider $b=0.7$ as an example 
to verify the above surmise and make an attempt to qualitatively analyze the influence of the superconducting pairing parameter $\Delta$ on the mobility edge. By 
taking four different $\Delta$, we plot the excitation spectra and \rm{IPR} as a function of $V$ under PBC, which are shown in Fig.~\ref{f4}(a), \ref{f4}(b), \ref{f4}(c), 
and \ref{f4}(d) respectively. According to the numerical results, the distinction between the extended states and the localized states can be readily seen from the \rm{IPR} (the color bar shows). 
The transition boundary in energy is just the mobility edge. When $\Delta$ is small, the low-energy excitation states are extended states, while those states with higher energy 
are localized. When $\Delta$ gets larger, the mobility edge moves towards the high-energy region. Moreover, we notice that when the quasiperiodic potential strength $V$ is small, all 
the excitation states are extended. Moreover, the extended region gets larger when $\Delta$ increases. In other words, the superconducting pairing is robust against the weak disorder 
and makes the system more extended. 
From Fig.~\ref{f4}(a), we also notice that when $V=2t$, the \rm{IPR} of the lowest excitation state 
approaches zero,  signaling the extended state. The result answers why the $\Psi$ and $\Psi$ in Fig.~\ref{f3}(d) and \ref{f3}(e) distribute throughout the whole system.

\section{summary}\label{S4}
Herin, a quasiperiodic $p$-wave superconducting model with coexistence of the homogeneous potential and quasiperiodic potential has been investigated. We proposed a 
modified real-space-Pfaffian method, which is beneficial and accurate to obtain the topological invariant of this system. We have demonstrated that the topological phases are protected by the gap. However, compared to the purely quasiperiodic case, the gap in the topological nontrivial is more wider. We have argued that this phenomenon occurs because the homogeneous potential has more impact on the gap. Besides, we have uncovered that there are mobility edges in the $p$-wave superconducting model. We have argued that the mobility edge originates from the competition between the extended $p$-wave pairing and the localized quasi-disorder. Furthermore, we have discussed the influence of the superconducting pairing parameter on the mobility edge. From the analysis, we have arrived at a qualitative fact that superconducting pairings tend to make system extended and stronger pairings move the mobility edge towards to high-energy region. 
In general, our theoretical work, i.e., the modified real-space Pfaffian method, overcomes the technical problem of using Pfaffian method to solve the topological invariants of a general 
$p$-wave superconductor and makes up for the limitation of the original Pfaffian. Moreover, the uncovered mobility edges promote a further understanding on the quasiperiodic $p$-wave superconductors. We note that a similar potential has been realized by the Raman coupling \cite{Raman} and the $p$-wave pairings can be induced by the superconducting proximity 
effect \cite{MF_3}. We expect that these intriguing quantum phenomena predicted in our study will be realized in near future experiments by means of these experimental techniques.   


\section{acknowledge}
Gao Xianlong and Shujie Cheng acknowledge the support from NSFC under Grants No.11835011 and No.11774316. Tong Liu acknowledges Natural Science Foundation of Jiangsu Province (Grant No. BK20200737) and NUPTSF (Grant No.NY220090 and No.NY220208).

\onecolumngrid
\begin{appendix}*
\section{Modified real-space-Pfaffian method}
As mentioned in the main text, we can perform a Schur decomposition \cite{schur} on a general anti-symmetric matrix $h$, i.e., $h=UDU^{T}$. After this operation, we obtain 
a $2L\times 2L$ anti-symmetric tridiagonal matrix $D$, which has the following configuration
\begin{equation}
D=\left(
\begin{array}{ccccccc}
0 & a_{1} & & & & & \\
-a_{1} & 0 & b_{1} & & & & \\
& -b_{1} & 0 & a_{2} &  & O &  \\
& & -a_{2} & \ddots & \ddots & & \\
& & & \ddots & 0 & b_{L-1} & \\
&O &  & & -b_{L-1} & 0 & a_{L} \\
& & & & & -a_{L} & 0 
\end{array}
\right).
\end{equation}
According to the original definition of the Pfaffian (Eq.~(\ref{Pf})), we can easily obtain the Pfaffian of the matrix $D$, which is given as 
\begin{equation}\label{Pf_D}
{\rm Pf}(D)=\prod \limits_{i=1}^L a_{i}. 
\end{equation}
Therefore, to obtain the Pfaffian of a general anti-symmetric matrix $h$, a standard strategy is that we first calculate to unitary matrix $U$ and the target matrix $D$ by 
the Schur decomposition, and then obtain the Pfaffian of $D$ by Eq.~(\ref{Pf_D}), and finally obtain the Pfaffian of $h$ by Eq.~(\ref{Pf_new}). We name the Pfaffian 
method after Schur decomposition as the modified real-space-Pfaffian method. In the following, we will show that how this method is effectively applied to obtain the 
topological phase diagram of $p$-wave superconducting models.

\begin{itemize}
  \item [1)] Test on the Kitaev model \cite{Kitaev}($V_{n}=V$). We consider $t=1$, $\alpha=(\sqrt{5}-1)/2$, $\Delta=0.5t$ and $L=5$ in all our tests. $V=1.5t$ and $V=2.5t$ 
  are two chosen parameters. We have known that the topological boundary is located at $V_{c}=2t$. Therefore, $V=1.5t$ corresponds to the topological non-trivial phase, and gives $M=-1$; 
  $V=2.5t$ corresponds to the topological non-trivial phase, and gives $M=1$; Taking $V=1.5t$ and other parameters into $h$, and performing the Schur decomposition, we have
  \begin{small}
  \begin{equation}
  U=\left(
  \begin{array}{cccccccccc}
  0 & -0.44721 & 0.62997 &  -0.019914 &   0.052389  &        0 &  -0.18026 &     0 &  0.60622 & 0 \\
   0.44721  & 0 & -0.019876  &-0.62876   &         0 &  0.065277   &         0 &  0.32194   &  0 &  0.54439 \\
   0  &-0.44721 & -0.54043 &  0.017083  & 0.32809 &    0 & -0.63226  &  0 &  0.015899 & 0 \\
   0.4721   & 0  & 0.014868 &  0.47033   &         0 & -0.42257 & 0 & -0.41826 &  0  &  0.47441 \\
   0 & -0.44721 &  0.24447 & -0.0077279 & -0.58324 &  0 & -0.21050 & 0 & -0.59640 & 0 \\
   0.44721   &0  & -0.0041805 & -0.13225  &          0 &  0.61846  & 0 & -0.58043 & 0 & -0.25119 \\
   0 & -0.44721 &  0.14487 & -0.0045795  & 0.61562 &  0 &  0.50216 & 0 & -0.38449 & 0 \\
   0.44721  & 0 & -0.0081034 & -0.25635  &          0 & -0.57812 &  0 &  0.059529 & 0 & -0.62965 \\
   0 & -0.44721 & -0.47888 &  0.015138 & -0.41285 & 0 &  0.52085 &  0 &  0.35877 & 0 \\
   0.44721  & 0  & 0.017292 &  0.54703  &          0 &  0.31695 &  0  &  0.61723  & 0 & -0.13796 
  \end{array}
  \right), 
  \end{equation}
  \end{small}
and 
\begin{small}
\begin{equation}
D=\left(
\begin{array}{cccccccccc}
0 & -0.50 &  0 & 0 &  0 & 0 &  0 & 0 & 0 & 0 \\
   0.50 & 0 &  0 & 0 &  0 &  0 &  0 &  0 & 0 & 0 \\
            0    &        0  & 0 & -3.1730  &  0 & 0 & 0 &  0 &  0 & 0 \\
            0     &       0  & 3.1730  & 0 &  0 &   0 &  0 &  0 &   0 & 0 \\
            0     &       0   &         0    &        0     &       0  &-3.1730  &    0 & 0 &   0  & 0\\
            0      &      0    &        0     &       0  & 3.1730 &            0  & 0 &  0  & 0 & 0\\
            0      &      0     &       0     &       0   &         0  &          0  &     0 &  1.2971 & 0  & 0\\
            0      &      0      &      0      &      0    &        0   &         0 & -1.2971 &    0  &  0  & 0\\
            0       &     0      &      0      &      0    &        0   &         0  &    0 &   0  &  0 & 1.2971 \\
            0       &     0      &      0      &      0    &        0   &         0   &  0 &  0 & -1.2971 & 0
\end{array}
\right).
\end{equation}
\end{small}
 It is easy to get $Det(U)=1$ and ${\rm Pf}(D)=-8.4688$. According to Eq.~(\ref{Pf_new}), we obtain the topological invariant $M=sgn({\rm Pf}(h))=-1$. Next, we take $V=2.5t$ and other 
 parameters into $h$. After performing the Schur decomposition, we have  
 \begin{small}
 \begin{equation}
 U=\left(
 \begin{array}{cccccccccc}
 -0.41412 & -0.16884 & -0.0048475 &  0.63189 & -0.026194 &  0.0015328 & -0.0071719 &  0.62251 &  0.11151 & 0 \\
  -0.16884 &  0.41412 & -0.62926 & -0.0048273 & -0.0036987  & -0.063207 &  0.5053 &  0.0058216  &    0 &   0.38031 \\
  -0.41412  & -0.16884 &  0.00404 & -0.52663 & -0.3496  &  0.020458 & -0.0034379 &  0.29841 & -0.55762 &  0   \\
  -0.16884 &  0.41412 &  0.47187 &  0.0036199 &  0.0246 &  0.42038 &  0.51782 &  0.0059658  &   0 &  -0.36308   \\
  -0.41412 & -0.16884 & -0.0016894 &  0.22022 &  0.59186 & -0.034635 &  0.0050471 & -0.43808 & -0.45614 &  0      \\
  -0.16884  & 0.41412 & -0.13424 & -0.0010298 & -0.036105 & -0.61699 & -0.18527 & -0.0021345  & 0 &  -0.60471  \\
  -0.41412 & -0.16884 & -0.0013065 &  0.17031 & -0.60805 &  0.035582 &  0.0065572 & -0.56916 &  0.27571 &  0  \\
  -0.16884 &  0.41412 & -0.25467 &  -0.0019537 &  0.033819 &  0.57793 & -0.63232 & -0.007285 &           0 & -0.010648  \\
  -0.41412 & -0.16884 &  0.0038034 &  -0.49579 &  0.39199 & -0.022938 & -0.00099451 &  0.0086322 &  0.62654 & 0  \\
  -0.16884 &  0.41412 &  0.5463 &  0.0041909 & -0.018615 & -0.31811 & -0.20553 & -0.0023678 & 0 &  0.59813 
 \end{array}
 \right), 
 \end{equation}
 \end{small}
 and 
 \begin{small}
 \begin{equation}
 D=\left(
 \begin{array}{cccccccccc}
  0 & -0.50 & 0 &  0 &   0 &    0 &    0 &    0 &  0 & 0 \\
   0.50   &         0  & 0 &   0 &  0 &   0 &   0 &   0 &   0  & 0 \\
            0  &          0   &         0  &  4.1598 &  0 & 0 &  0 &   0 &  0  & 0 \\
            0   &         0   & -4.1598  &           0 &  0 &   0 &   0 &   0 &  0  & 0 \\
            0  &          0    &        0   &         0  & 0 &  -4.1598 &   0 &   0 &    0  & 0 \\
            0  &          0    &        0   &         0   & 4.1598 &  0 &    0 &   0 &    0 & 0 \\
            0   &         0    &        0   &         0   &         0    &        0   &         0  & -2.1086  &  0  & 0 \\
            0    &        0     &       0    &        0    &        0     &       0 &  2.1086 &            0  & 0 & 0 \\
            0    &        0     &       0    &        0    &        0     &       0   &         0    &        0    &        0 & 2.1086 \\
            0   &         0     &       0    &        0    &        0     &       0   &         0     &       0  & -2.1086  & 0 
 \end{array}
 \right).
 \end{equation}
 \end{small}
 In this case, we have $Det(U)=-1$ and ${\rm Pf}(D)=-3.8469$. According to Eq.~(\ref{Pf_new}), we get $M=1$. If leaving the $V$ changing while other parameters 
 invariable, by means of this method, we can obtain the topological phase diagram of the Kitaev model, as shown in Fig.~\ref{f1_appen}.
 \begin{figure}[H]
\centering
\includegraphics[width=0.5\textwidth]{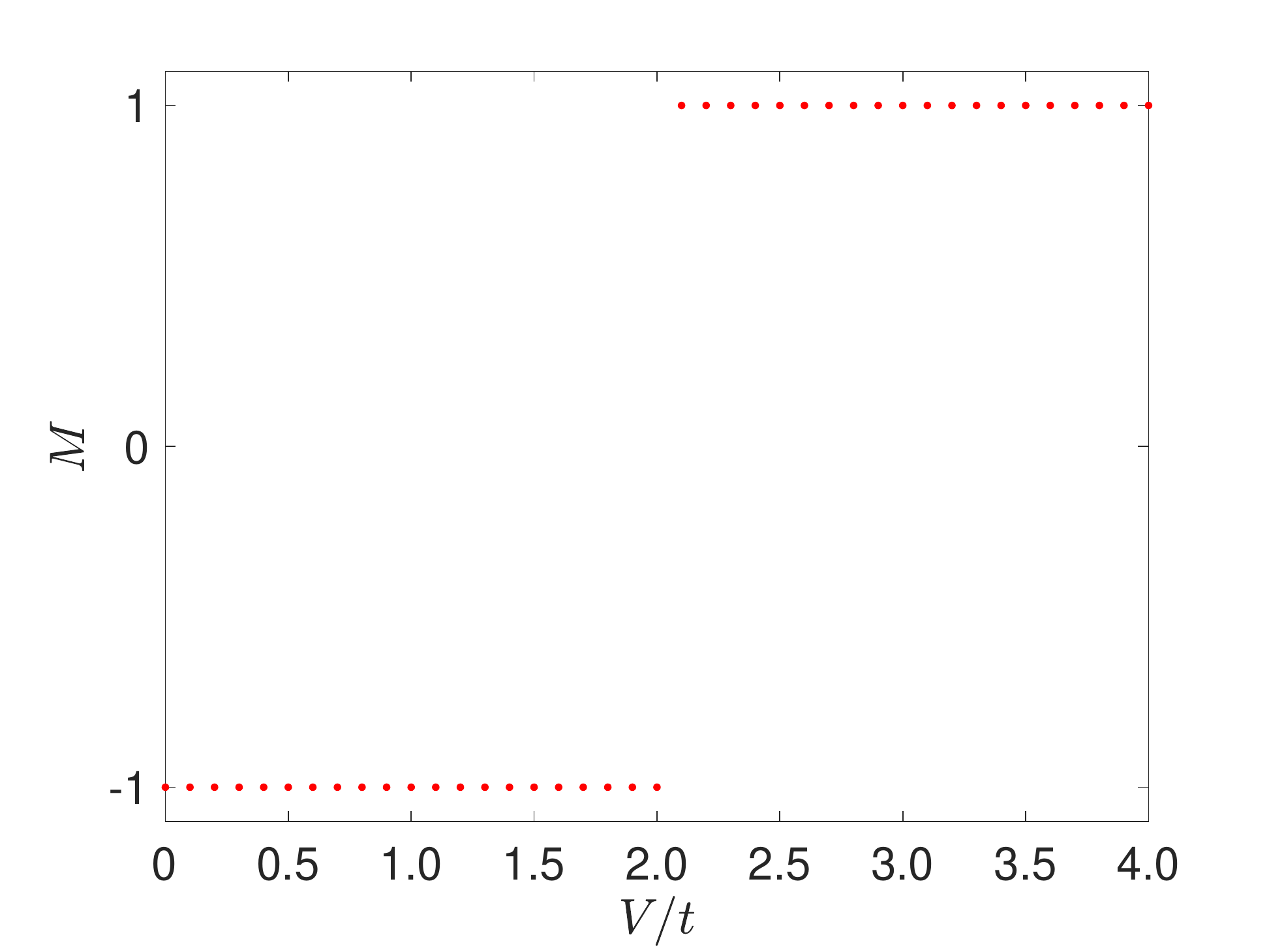}\\
\caption{(Color Online) $M-V/t$ topological phase diagram of the Kitaev model.}\label{f1_appen}
\end{figure}

  \item [2)] Test on the quasiperiodic $p$-wave superconducting model \cite{cai, Wang}. This model requires $V_{n}=V\cos(2\pi\alpha {n})$, and it topological 
  boundary is located at $V_{c}=2t+2\Delta$. Without loss of generality, we take $\Delta=0.5t$, thus the topological boundary is $V_{c}=3t$.  $V=2.8t$ and $V=3.2t$ 
  are two chosen parameters, corresponding to the topological non-trivial phase ($M=-1$) and topological trivial phase ($M=1$) respectively. We first test the case 
  with $V=2.8t$. After performing the Schur decomposition, we get 
  \begin{small} 
  \begin{equation}
  U=\left(
  \begin{array}{cccccccccc}
  -0.10292 & -0.086979 & -0.014115 & -0.75278 & -0.048208 & -0.38713 &  0.077204 &  0.14370 &  0.48596 & \clap{-1.8239e-04} \\
  -0.024698 &  0.029224 &  0.60102 & -0.011270 & -0.45909 &  0.057169 & -0.57104 &  0.30680 &  \clap{2.0348e-05} &  0.054217 \\
   0.057680 &  0.048749 & \clap{-6.3534e-03} & -0.33884 & -0.012153 & -0.097596 &  0.25378 &  0.47234 & -0.76306 & \clap{2.8639e-04} \\
   0.041361 & -0.048939 &  0.027242 & \clap{-5.1081e-04} & -0.55188 &  0.068724 &  0.31347 & -0.16842 & \clap{-2.8067e-04} & -0.74782 \\
  -0.071450 &  -0.060387 &  \clap{2.61042-03} &  0.13922 & -0.10006 & -0.80349 & -0.22492 & -0.41864 & -0.30068 & \clap{1.1285e-04} \\
  -0.24982 &  0.29559 &  0.31525 & \clap{-5.9112e-03} &  0.62080 & -0.077306 & -0.15776 &  0.084763 & \clap{-2.1473e-04} & -0.57214 \\
   0.59525 &  0.50308 & \clap{-7.1582e-03} & -0.38176 &  0.018809 &  0.15104 & -0.21896 & -0.40753 & -0.097991 & \clap{3.6778e-05} \\
  -0.54896 &  0.64954 &  0.12755 & \clap{-2.3916e-03} & -0.23261 &  0.028966 &  0.31620 & -0.16989 &  \clap{1.0393e-04} & 0.27691 \\ 
   0.45824 &  0.38728 &  \clap{7.3351e-03} &  0.39119 & -0.049331 & -0.39615 &  0.23475 &  0.43692  & 0.28560 & \clap{-1.0719e-04} \\
   0.22490 & -0.26611 &  0.72251 & -0.013548 &  0.17289 & -0.021530 &  0.47621 & -0.25586 &  \clap{6.9004e-05} & 0.18386 
 \end{array} 
   \right),
  \end{equation}
  \end{small}
  and 
  \begin{small}
  \begin{equation}
  D=\left(
  \begin{array}{cccccccccc}
  0 & -3.5300 &  0 &  0 &  0 &   0 &  0 &   0 &   0 &   0 \\
   3.5300 &  0 &   0 &  0 &  0 &  0 &  0 &   0 &  0 &   0 \\
            0      &      0    &   0 &  -3.1062 &   0 &   0 &   0 &   0 &  0  &  0 \\
            0      &      0    &  3.1062  &   0 &  0 &   0 &  0 &   0 &   0 &  0 \\
            0       &     0     &       0    &        0  & 0 &   2.4913 &   0 &  0 &  0 &   0 \\
            0      &      0     &       0     &       0   &  -2.4913   &  0 &   0 &  0 &   0 &   0 \\
            0      &      0     &       0    &        0   &         0      &       0 &  0 &  -2.1429 &  0 & 0 \\
            0      &      0     &       0     &       0    &        0      &      0  & 2.1429 &   0 &   0 &  0 \\
            0      &      0     &       0    &        0    &        0      &      0   &         0  &          0  &   0 &   -0.028422 \\
            0      &      0     &       0     &       0    &        0      &      0   &         0   &         0  &  0.028422 &  0
  \end{array}
  \right).
  \end{equation}
  \end{small}
 In this case, we have $Det(U)=-1$ and ${\rm Pf}(D)=1.6637$. Therefore, the topological invariant is $M=sgn(Det(U){\rm Pf}(D))=-1$. Next, we take $V=3.2t$ and other 
 parameters into $h$. After performing the Schur decomposition, we obtain
 \begin{small}
 \begin{equation}
 U=\left(
 \begin{array}{cccccccccc}
 0.011327 & -0.13876 &  0.10555 &  0.73415 & -0.42287 & -0.076369 & -0.14586 & -0.11929 &  0.45864 &  \clap{1.0053e-04} \\
   0.012075 &  \clap{9.8569e-04} & -0.58865 &  0.084633 & -0.094622 &  0.52394 &  0.37923 & -0.46370 &  \clap{1.3679e-05} & -0.062405 \\
  -0.0045091 &  0.055237 &  0.043898 &  0.30532 & -0.12858 & -0.023222 & -0.37712 & -0.30842 & -0.80459 & \clap{-1.7636e-04} \\
  -0.057731 & -0.0047127 & -0.020556 &  0.0029554 & -0.087289 &  0.48334 &  -0.23696 &  0.28974 & \clap{-1.7188e-04} &  0.78414 \\
   0.0075514 & -0.092506 & -0.020357 & -0.14159 & -0.74798 & -0.13508 &  0.43505 &  0.35580 & -0.27810 & \clap{-6.0959e-05} \\
   0.35934 &  0.029334 & -0.30339  & 0.043620 &  0.11036 & -0.61106 &  0.18838 & -0.23034 & \clap{-1.2045e-04} &  0.54949 \\
  -0.064434 &  0.78933 &  0.056941 &  0.39604 &  0.16551 &  0.029890 &  0.32728 &  0.26766 & -0.075370 & \clap{-1.6521e-05} \\
   0.85592 &  0.069870 & -0.16543 &  0.023785 & -0.033619 &  0.18615 & -0.23825 &  0.29132 &  \clap{5.2414e-05} & -0.23912 \\
  -0.047563 &  0.58265 & -0.059396 & -0.41311 & -0.43149 & -0.077926 & -0.37328 & -0.30528 &  0.24346 &  \clap{5.3365e-05} \\
  -0.35803 & -0.029226 & -0.71653 &  0.10302 &  0.041274 & -0.22854 & -0.32950 &  0.40290 &  \clap{3.2591e-05} & -0.14868
 \end{array}
 \right),
 \end{equation}
 \end{small}
  and 
  \begin{small}
  \begin{equation}
  D=\left(
  \begin{array}{cccccccccc}
  0 &  3.8730 &   0 & 0 & 0 & 0 & 0 & 0 & 0 & 0  \\
  -3.8730 &  0  & 0 & 0  & 0 & 0  & 0 & 0 & 0 &  0 \\
            0  &          0  & 0 & -3.3699 &  0 & 0 & 0 &  0 &  0 &  0 \\
            0  &          0  & 3.3699  & 0  & 0 &  0 &  0 & 0 & 0 &  0 \\
            0   &         0    &        0     &       0 &  0 & 2.6843 & 0 & 0 &  0 &  0  \\
            0  &          0    &        0      &      0  & -2.6843 & 0  & 0 & 0 & 0 &  0 \\
            0  &          0    &        0      &      0   &         0      &   0 & 0 & -2.3647 & 0 &  0 \\
            0  &          0     &       0      &      0    &        0      &      0  &  2.3647 &  0 &  0 &  0 \\
            0  &          0     &       0      &      0    &        0       &     0     &     0     &    0 &  0 & -0.047520 \\
            0  &          0    &        0      &      0    &        0       &     0     &     0     &    0 &  0.047520 &  0  
  \end{array}
  \right).
  \end{equation}
  \end{small}
  Along the same strategy, we have $Det(U)=-1$ and ${\rm Pf}(D)=-3.9369$, so the topological invariant is $M=sgn(Det(U){\rm Pf}(D))=1$. If we change the 
  $V$ continuously and keep other parameters invariable, then we can obtain the topological phase diagram, as shown in Fig.~\ref{f2_appen}.
\begin{figure}[H]
\centering
\includegraphics[width=0.5\textwidth]{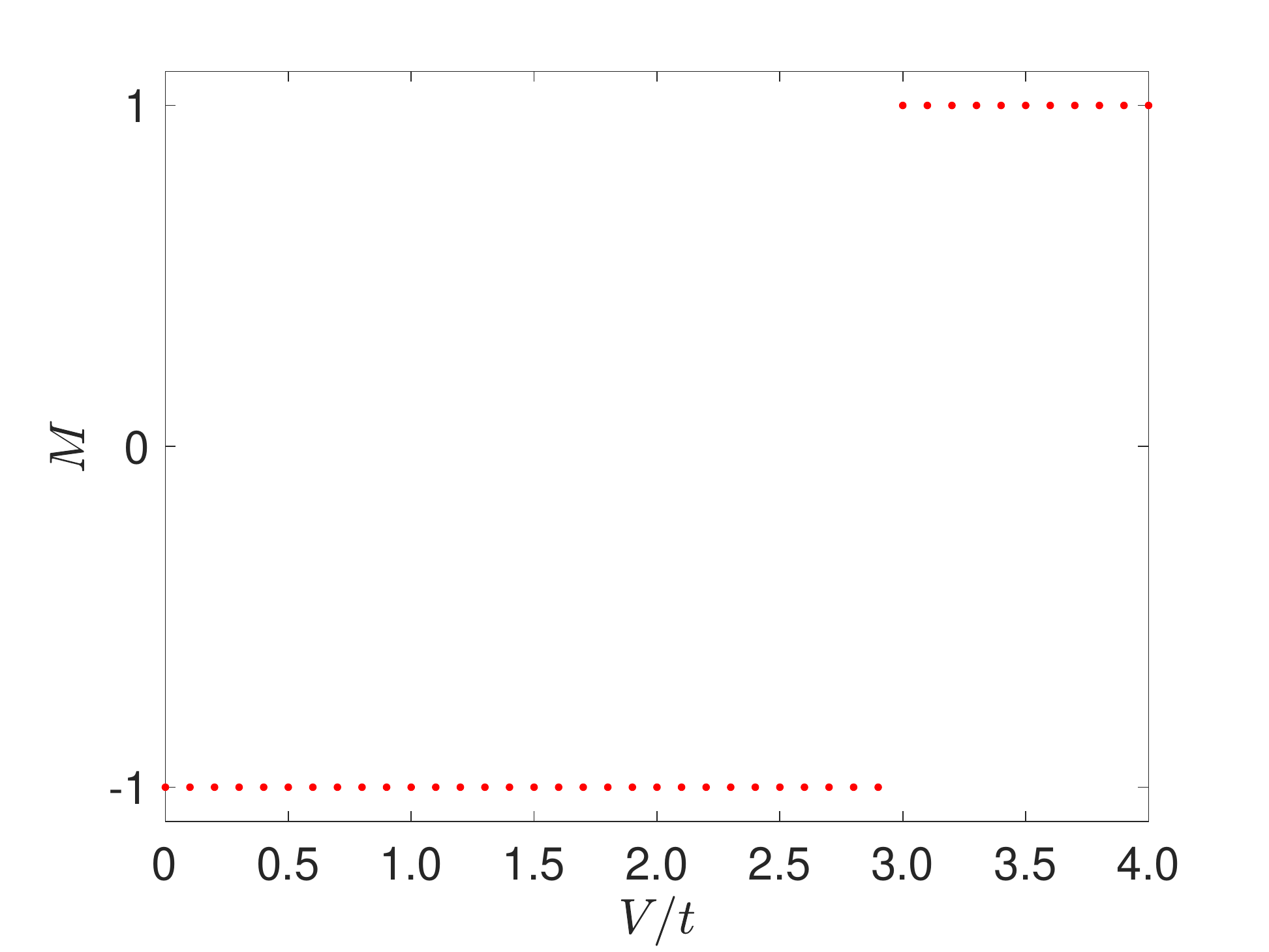}\\
\caption{(Color Online) $M-V/t$ topological phase diagram of the quasiperiodic $p$-wave superconducting model.}\label{f2_appen}
\end{figure}

  \item [3)] Test on our model (see Eq.~(\ref{eq1})). The potential has been presented in Eq.~(\ref{potential}). We consider $b=0.95$ in our test. According to the topological phase 
  diagram in Fig.~(\ref{f1}), we know the numerical topological boundary is located at about $V_{c}=1.5t$. $V=1.2t$ and $V=1.7t$ are two chosen parameters, corresponding to the topological 
  non-trivial phase and topological trivial phase respectively. We first test the case with $V=1.2t$. After performing the Schur decomposition, we have 
  \begin{small}
  \begin{equation}
  U=\left(
  \begin{array}{cccccccccc}
  -0.023316 & -0.23546 & -0.094225 &  0.063506 &  0.31829 &  0.046668 &  0.85945  ~~~~& \clap{4.9924e-04} & -0.29822 & \clap{-5.4378e-03} \\
  -0.10571 &  0.010468 & -0.092276 & -0.13691 &  0.11105 & -0.75736 & \clap{-1.6668e-04}~~~~ &  0.28693 & \clap{-9.8719e-03} &  0.54139 \\
   \clap{2.4810e-03} &  0.025054 &  0.28855 & -0.19447 & -0.83687 & -0.12270 &  0.39619 ~~~~&  \clap{2.3014e-04} ~~~~&  0.076820 &  \clap{1.4008e-03} \\
   0.016300 & \clap{-1.6141e-03} &  0.26911 &  0.39929 & -0.069709 &  0.47543 & \clap{-2.0030e-04} ~~~~&  0.34482 & -0.011788 &  0.64649 \\
   \clap{2.9746e-03} &  0.030039 & -0.71121 &  0.47934 & -0.24651 & -0.036144 &  0.13654 ~~~~~&  \clap{7.9316e-05} ~~~~&  0.42745 &  \clap{7.7944e-03} \\
   0.035612 & \clap{-3.5265e-03} & -0.47172 & -0.69990 & -0.058984 &  0.40229 & \clap{-1.7914e-04} ~~~~&  0.30839 & \clap{-2.9356e-03} &  0.16099 \\
  -0.010398 & -0.10500 &  0.29848 & -0.20117 &  0.31062 &  0.045544 &  0.19598 ~~~~&  \clap{1.1384e-04}~~ ~~&  0.84976 &  0.015495 \\
  -0.23504 &  0.023275 &  0.090986 &  0.13500 &  \clap{8.1582e-03} & -0.055641 ~~~& \clap{-4.7133e-04} ~~~~&  0.81139 &  \clap{9.2280e-03} & -0.50607 \\
   0.095100 &  0.96036  & 0.024252 & -0.016345 &  0.14154 &  0.020753  & 0.21754 ~~&  \clap{1.2636e-04}~~ ~~&  \clap{4.4215e-03} &  \clap{8.0623e-05} \\
   0.96038 & -0.095103 &  0.025035 &  0.037146 &  0.017590 & -0.11997~~~ & \clap{-1.2366e-04} ~~~~&  0.21288 &  \clap{1.4808e-03} & -0.081207 \\
  \end{array}
  \right),
  \end{equation}
  \end{small}
  and 
  \begin{small}
  \begin{equation}
  D=\left(
  \begin{array}{cccccccccc}
  0 & -6.4697 & 0 &  0 &  0 &  0 &  0  & 0 & 0 &  0 \\
   6.4697 & 0 & 0  & 0 & 0  & 0 & 0  & 0 & 0 &  0 \\
            0     &       0 & 0  & 3.7355 & 0 &  0 &  0 &  0 & 0 & 0 \\
            0     &       0 & -3.7355 & 0 & 0 & 0 &  0 & 0 & 0 &  0 \\
            0     &       0  &          0   &         0     &       0 & -1.8606 & 0 &  0 & 0 & 0 \\
            0     &       0  &          0   &         0  & 1.8606     &       0 &  0 &  0 &  0 & 0 \\
            0     &       0   &         0   &         0    &        0    &        0 & 0 & -0.33654  & 0 &  0 \\
            0     &       0   &         0   &         0    &        0    &        0 &  0.33654 & 0 &  0 &  0 \\
            0     &       0   &         0   &         0    &        0    &        0   &         0    &        0     &        0 &  -0.60564 \\
            0     &       0   &         0   &         0    &        0    &        0   &         0     &       0  & 0.60564  &          0
  \end{array}
  \right).
  \end{equation}
  \end{small}
  Easily, we have $Det(U)=-1$ and ${\rm Pf}(D)=9.1651$. Therefore, the topological invariant is $M=sgn(Det(U){\rm Pf}(D))=-1$. Next, we test the case with 
  $V=1.7t$. In the same way, we perform the Schur decomposition, then we get 
  \begin{small}
  \begin{equation}
  U=\left(
  \begin{array}{cccccccccc}
  \clap{-2.6224e-07} & 0.17518  & 0.056526 & -0.050351 & -0.37571 & -0.033873 & -0.52435 &  \clap{2.0649e-05} & ~ 0.73915 ~~&~~ \clap{9.0440e-04} \\
  -0.077407 & \clap{-1.1588e-07} & -0.072657 & -0.081567 &  0.062177 &  -0.68965 & \clap{-1.8448e-05} & -0.46847 &~ \clap{-6.5106e-04} ~~&~~  0.53210 \\
   \clap{2.1014e-08} & -0.014037 & -0.21481 &   0.19135 &   0.85794 &   0.077350 &  -0.090087 &   \clap{3.5476e-06}  & ~ 0.40852 ~~~&   \clap{4.9985e-04} \\
   0.0099040 &   \clap{1.4826e-08} &  0.27022 &   0.30336 &  -0.055726 &   0.61809 &  \clap{-1.6487e-05} & -0.41867 &~ \clap{-6.4098e-04}  & 0.52385 \\
   \clap{2.5680e-08} & -0.017155 &   0.67495 &  -0.60122 &   0.22258 &   0.020067  & 0.28462 & \clap{-1.1208e-05} &  ~0.22745~~~~  & \clap{2.7831e-04} \\
   0.019830 &  \clap{2.9685e-08}  & -0.59529 & -0.66829 & -0.031601  & 0.35050  & \clap{-1.6615e-07} & -0.0042191 & ~ \clap{-3.3454e-04}~~~~ &  0.27341 \\
  \clap{-1.1364e-07}  & 0.075912 & -0.22895 &  0.20394 & -0.23432 & -0.021126 &  0.79666 & \clap{-3.1372e-05} & ~ 0.45848 ~~& ~~ \clap{5.6099e-04} \\
  -0.17437 & \clap{-2.6104e-07} &  0.096535 &  0.10837 &  0.0063677 & -0.070628  & \clap{3.0359e-05} &  0.77094 & ~\clap{-7.2301e-04} ~~~ & 0.59089 \\
   \clap{1.4691e-06} & -0.98136 & -0.016346 &  0.014560 & -0.10135 & -0.0091379 & -0.035661 &  \clap{1.4043e-06} &  ~0.15759 ~~~&  \clap{1.9282e-04} \\
   0.98138  & \clap{1.4691e-06}  & 0.020723  & 0.023265 &  0.0072366 & -0.080266 &  \clap{4.1090e-06} &  0.10434 &~ \clap{-1.6659e-04}~~~~ &  0.13615
  \end{array}
  \right),
  \end{equation}
  \end{small}
  and 
  \begin{small}
  \begin{equation}
  D=\left(
  \begin{array}{cccccccccc}
  0 &  8.8733 &  0 & 0 &  0 &  0 & 0  & 0  & 0 &  0 \\
  -8.87330 &  0 & 0 & 0  & 0 & 0 &  0 &  0 &  0 & 0 \\
            0     &       0     &       0  &       -4.7433 &  0 & 0 &  0 & 0 &  0 & 0 \\
            0     &       0      &    4.7433 &           0 &  0 &  0 &  0 &  0 &  0 &  0 \\
            0     &       0      &      0       &     0 & 0  & 2.3372 &  0 &  0 & 0 &  0 \\
            0     &       0      &      0       &     0 & -2.3372 & 0 & 0 &  0 &  0 &  0 \\
            0     &       0      &      0       &     0  &          0     &       0  & 0 & 0.79243 &   0 & 0 \\
            0     &       0      &      0       &     0   &         0     &       0  &  -0.79243 &  0 & 0 & 0 \\
            0     &       0      &      0       &     0   &         0      &      0    &        0    &        0  & 0 &  0.089008 \\
            0     &       0      &      0       &     0   &         0     &       0    &        0    &        0  & -0.089008 &  0
  \end{array}
  \right).
  \end{equation}
  \end{small}
  After calculation, we have $Det(U)=-1$ and ${\rm Pf}(D)=-6.9383$. Therefore, in this case, the topological invariant is $M=sgn(Det(U){\rm Pf}(D))=1$. 
  If we change $V$ continuously and leave other parameters invariable, then we can obtain the topological phase diagram, as shown in Fig.~\ref{f3_appen}.
  \begin{figure}[H]
\centering
\includegraphics[width=0.5\textwidth]{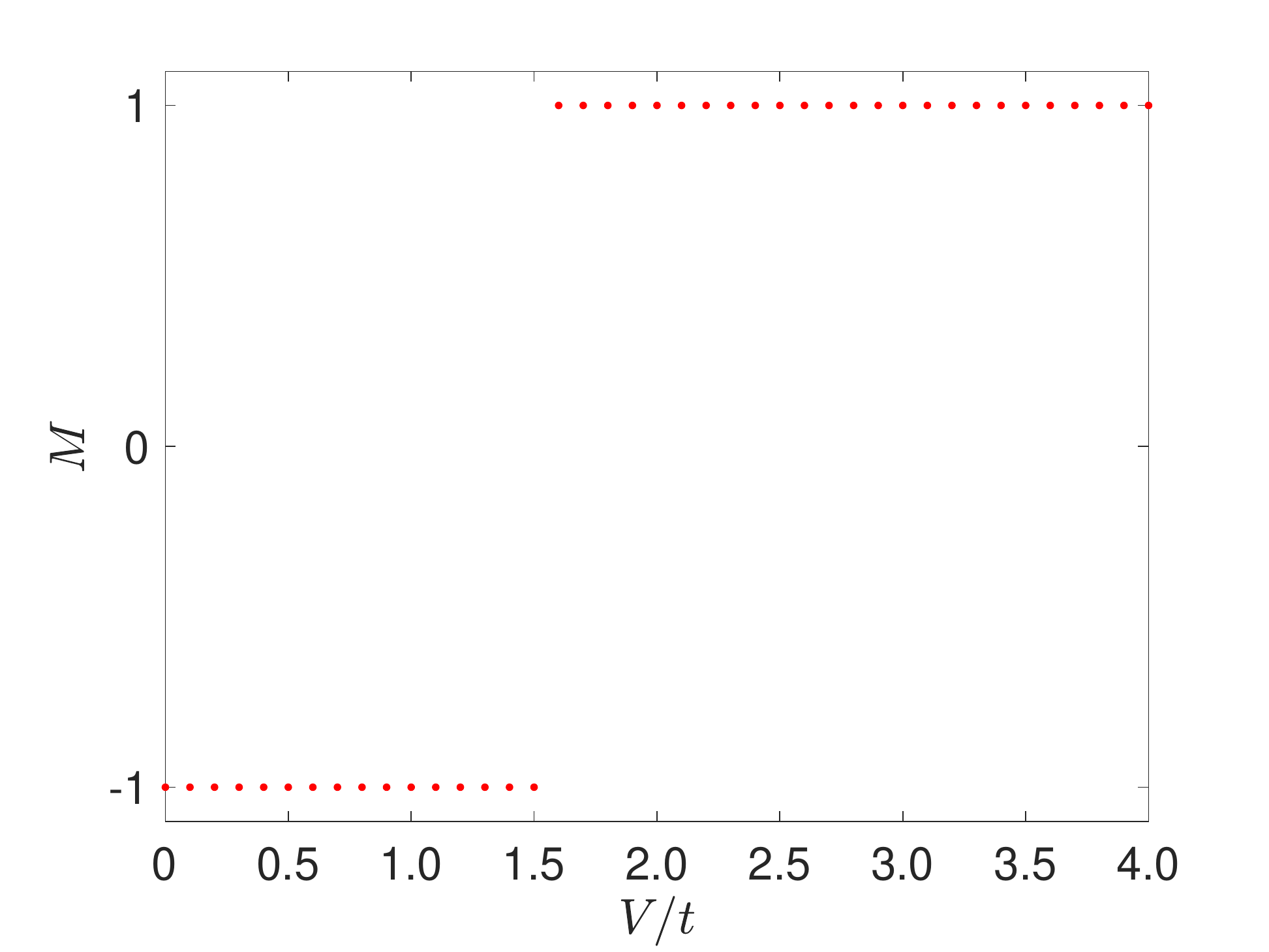}\\
\caption{(Color Online) $M-V/t$ topological phase diagram of our model.}\label{f3_appen}
\end{figure} 
  
\end{itemize}
From the above tests, it is readily seen that this modified real-space-Pfaffian method is very convenient and accurate to numerically obtain the 
topological phase diagram of $p$-wave superconducting model.

\end{appendix}

\twocolumngrid


\end{document}